\documentclass[aps,showpacs,epsf,twocolumn,prl]{revtex4}
\usepackage{amssymb}
\usepackage{amsmath}
\usepackage{graphicx,psfrag}
\usepackage{braket}
\usepackage{float}
\usepackage{subfig}
\usepackage{tikz}
\usepackage{epstopdf}
\usepackage{pgfplots}
\usepackage[colorlinks=true, citecolor=blue, urlcolor = blue, linkcolor= red, bookmarks=true]{hyperref}

\begin{document}

\def \beq{\begin{equation}}
\def \eeq{\end{equation}}
\def \bse{\begin{subequations}}
\def \ese{\end{subequations}}
\def \bea{\begin{eqnarray}}
\def \eea{\end{eqnarray}}
\def \bem{\begin{displaymath}}
\def \eem{\end{displaymath}}
\def \bem{\begin{bmatrix}}
\def \eem{\end{bmatrix}}
\def \Ps{\hat{\Psi}(\boldsymbol{r})}
\def \Pds{\hat{\Psi}^{\dagger}(\boldsymbol{r})}
\def \i{{\int}d^2{\bf r}}
\def \bl{\bar{\boldsymbol{l}}}
\def \c{\hat{c}_{n,m}}
\def \cp{\hat{c}_{n',m'}}
\def \cd{\hat{c}_{n,m}^{\dagger}}
\def \cdp{\hat{c}_{n',m'}^{\dagger}}
\def \bb{\bibitem}
\def \nn{\nonumber}

\def \bs{\boldsymbol}
\def \hkx{\hat{k}_{x}}
\def \hky{\hat{k}_{y}}
\def \bq{\bar{q_{y}}}

\def \bc{\begin{center}}
\def \ec{\end{center}}

\title{\textbf{A magnetic Hofstadter butterfly and its topologically quantized Hall conductance}}
\author{ Manisha Arora and Sankalpa Ghosh}
\affiliation{Department of Physics, Indian Institute of Technology Delhi, New Delhi-110016, India}

\begin{abstract}
The energy spectrum of massless Dirac fermions in graphene under two dimensional periodic magnetic modulation having square lattice symmetry is calculated. 
We show that the translation symmetry of the problem  is similar to that of the Hofstadter or TKNN problem and in the weak field limit the tight binding energy eigenvalue equation is indeed given by Harper Hofstadter hamiltonian. We show that due to its magnetic translational symmetry the Hall conductivity can be identified as a topological invariant and hence quantized. 
We thus extend the idea of Quantum Hall Effect to magnetically modulated two dimensional electron system. Finally we indicate possible experimental systems where this may be verified.  
\end{abstract}

\maketitle

\newpage
The Integer Quantum Hall effect (IQHE), one of the most fascinating phenomena in condensed matter physics and beyond, shows extremely precise 
quantization of magneto-transport (Hall conductivity) in  a two dimensional electron gas (2DEG) placed in a transverse magnetic field. It was first discovered in a system with non-relativistic dispersion 
 \cite{Klitzing, Tsui, Laughlin1, Halperin}  (Silicon MOSFET and semiconductor heterostructure), and subsequently in a 2DEG with relativistic dispersion  \cite{Novoselov1, Kim}  realised  in Graphene.
 In a seminal paper Thouless, Kohmoto, Nightingale, and den Nijs (TKNN) \cite{Thouless1} identified this quantized Hall conductance of a 2DEG in a transverse magnetic field and a strong periodic potential \cite{Hofstadter}
with a topological invariant \cite{Avron1} called first Chern number defined in the space of Bloch vectors. This remarkable idea not only  explained the robust quantization of a transport quantity, but fundamentally modified the conventional understanding of band structure of transport. 
This idea was further significantly generalised when  Haldane \cite{Haldane} showed that such  topological quantization of Hall conductivity can be achieved for a fully filled band even in the absence of a net magnetic field leading to anomalous quantum Hall Effect (AQHE). 
Haldane's seminal work was further generalised in systems without any external magnetic field,  that respect time reversal symmetry \cite{BHZ, KM} and lead to the discovery of present day Topological Insulators \cite{HK}.

This letter considers two dimensional gas of massless Dirac fermions with relativistic dispersion, the charge carriers in monolayer graphene, in the presence of a periodically modulated transverse magnetic field.
We show that such magnetic field is composed of two parts. One part corresponds to a uniform field, like that in a prototype quantum Hall system. The other part gives a periodic modulation which has zero net flux through each unit cell. Thus the second part  gives 
qualitatively similar flux condition  as in the Haldane's problem\cite{Haldane} leading to AQHE, but realised in a different  lattice geometry as compared to Haldane' s construction. 
Subsequently we write the resulting vector potential as a combination of the usual symmetric gauge vector potential 
for a uniform transverse field and a periodic vector potential for the modulated part, whose form we also explicitly construct. This decomposition enables us to identify that the magnetic translation operator \cite{Zak, Fischbeck, Florek} for this problem is same as that in the TKNN \cite{Thouless1} problem. 

This has interesting consequences. 
The corresponding hamiltonian is a perturbed form of the hamiltonian used in the classic Hofstadter problem \cite{Hofstadter}. However 
here the periodicity comes entirely from magnetic modulation. We therefore address this as a magnetic Hofstadter problem. In tight-binding approximation and  in the weak field limit we show that the problem is indeed Hofstadter-Harper equation.  We analyse its spectrum. 
Finally the properties of the magnetic Bloch functions are used to show that the Hall conductivity stays quantised as in the TKNN problem and can be identified with the Chern number of a filled band. Thus our decomposition of magnetic field 
profile allows us to show the topological quantization of Hall conductivity for a generic magnetically modulated 2DEG. 

The effect of periodic magnetic modulation in one  \cite{Peeters, Krakovsky}
and two dimension \cite{Chang, Oh} on a 2DEG  has been considered earlier in a number of papers. They mostly explore the resulting band structure \cite{Peeters, Krakovsky} and the perturbation of 
Landau levels formed in a uniform magnetic field. 
Our decomposition of a general periodic magnetic modulation and the identification of the magnetic translation symmetry not only provides
a unifying theoretical framework to these previous studies but also establishes a connection between conventional 
Quantum Hall effect and AQHE espoused in Haldane like models. We conclude the paper by indicating some physical systems where this prediction can be put to experimental test.

\begin{figure}
\includegraphics[width=7cm,height=7cm,keepaspectratio]{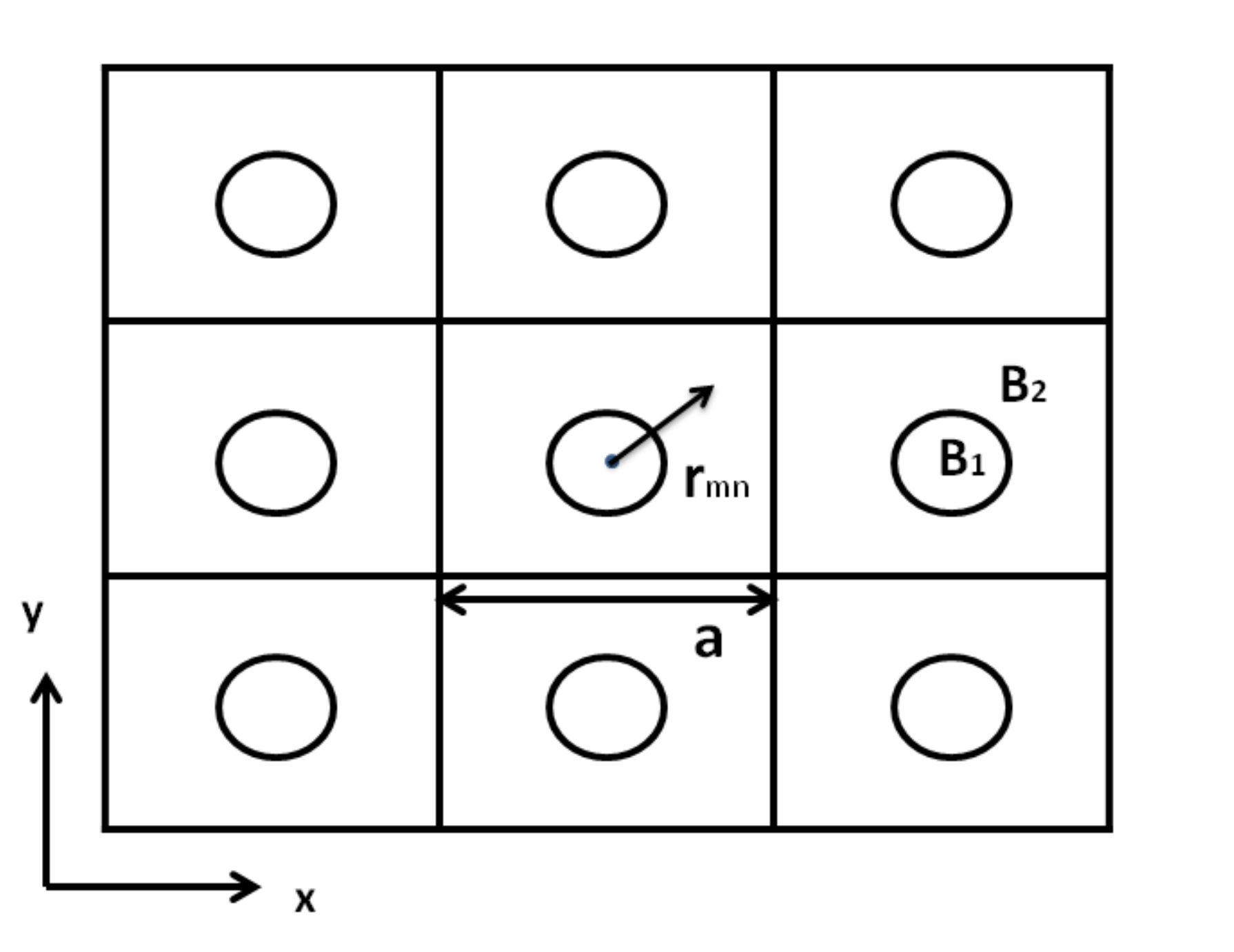}
\caption{Monolayer Graphene under a two dimensional magnetic field array with inside circular regions with radius $R_{s}$ having $B_{1}\bs{\hat{z}}$ and that of outside having $B_{2}\bs{\hat{z}}$}
\label{Fig1}
\end{figure}

We consider the charge carriers in monolayer graphene in a  transverse periodic magnetic field modulation (Fig. \ref{Fig1}) given by 
\beq B \bs{\hat{z}}= \Big[ B_{2}  + (B_{1} - B_{2}) \sum_{m,n} \Theta (R^{s} - r_{mn}) \Big] \bs{\hat{z}} \label{B}\eeq 
in the $\bs{k}\cdot \bs{p}$ representation, described by the massless dirac hamiltonian  
\beq 
\hat{H}=v_{F} (\bs{\sigma} \cdot \bs{\Pi})  \label{Ham1}
\eeq
where
$\bs{\Pi} =\bs{p}+\frac{e\bs{A}}{c}$ is the canonical momentum operator 
with $\bs{A}$ as the vector potential for field (\ref{B}).  $R^{s}$ is the radius of the circular region within each unit cell, 
$\bs{r}_{mn} = \bs{r} - \bs{R}_{mn}$
with $\bs{r}  =  x\bs{\hat{x}} + y\bs{\hat{y}}$.  The lattice vector $\bs{R}_{mn}  = ma \bs{\hat{x}} + na\bs{\hat{y}}$
and 
\bea 
\Theta (R^{s} - r_{mn})
& = & 
\begin{cases}
1, ~~
\text{if}~\ r_{mn} \leq R^{s}, \\
0, 
 ~~\text{if}~\ r_{mn} > R^{s} \end{cases}   \eea 
with 
$m,n \in I (-N_{max}, +N_{max})$ and $a$ is the lattice constant.  The step function modelling can be justified by assuming that the magnetic field gradient at the step is much 
sharper than the typical Fermi-wavelength of the electron \cite{Egger, Peeters, Alain, Neetu}. 
Such two dimensional periodic modulation corresponds to that of a square lattice and can be made using nanolithographic techniques \cite{expt1, Bader}. 
Using Fourier theorem, the periodic magnetic field profile given by Eq. $(\ref{B})$ can be written as 
\beq
\bs{B}=\sum_{\bs{G}} \bs{B}_{\bs{G}} e^ {i \bs{G}.\bs{r}} =\bs{B}_{0} + \sum_{\bs{G} \neq 0} \bs{B}_{\bs{G}} e^ {i \bs{G}.\bs{r}} \label{Fourier}
\eeq
Here $\bs{G}$ is the reciprocal lattice vector of the periodic lattice, $\bs{B}_{0}=\bs{B}_{u}$ is the uniform magnetic field and is given by spatial average of the field given in Eqn. $(\ref{B})$. The residual periodic field is defined as $\bs{B}_{p}= \sum_{\bs{G}\neq 0} \bs{B}_{\bs{G}} e^ {i \bs{G}.\bs{r}} $ that satisfies $
\bs{\nabla} \cdot \bs{B}_{p}=0$ with net flux through the unit cell due to $\bs{B}_{p}$ is zero (for details see  \cite{suppli}).
This decomposition remains valid even for an arbitrary periodic modulation.  Consequently if we apply uniform 
magnetic field $-\bs{B}_{u}$ to this system, under suitable condition this can realise Haldane like model \cite{Bruno, Snyman}. 
Now, following \cite{Brownbook} it can be shown that there corresponds a periodic vector potential for the residual periodic field 
$\bs{B}_{p}$. Accordingly we can decompose the magnetic field  and corresponding vector potential in uniform and periodic parts like 
\beq B \bs{\hat{z}} =  B_{u} \bs{\hat{z}} + B_{p} \bs{\hat{z}}~ \text{and}~ \bs{A}(\bs{r})  =   \bs{A}_u(\bs{r})+\bs{A}_{p}(\bs{r}) \nn
\eeq
with 
\bea B_{u}  & = &  \Big[ B_{2} + \frac{(B_{1}-B_{2}) \pi (R^s)^2}{a^2} \Big]  \nonumber \\
B_{p} & = & \Big[ \sum_{m,n}  ((B_{1} - B_{2})  \theta(R^{s}-r_{mn})) \nn\\
&-&\frac{(B_{1}-B_{2}) \pi (R^s)^2}{a^2} \Big]  \label{Bp}
 \label{magfield1} \eea
 \bea
 \bs{A}_{u}(\bs{r}) & = & \frac{1}{2} \bs{B}_{u} \times \bs{r}, ~ \text{Symmetric Gauge}  \nn \\
 \bs{A}_{p} (\bs{r}) & = & \frac{1}{2} \sum_{m,n} \Big[(B_{p}^{mn} \bs{\hat{z}}) \times (r_{mn} \bs{\hat{r}}_{mn})\Big] \label{Ap} \eea
where
\begin{equation}
B_{p}^{mn}=
\begin{cases}
\Big[ (B_1 - B_2) - \frac{(B_1-B_2) \pi (R^s) ^{2}}{a ^2} \Big] \Theta(R^{s}-r_{mn}), \\
\text{if}\ r_{mn} \leq R^{s} \\\\
(B_{2}-B_{1}) \frac{\pi (R^s)^2}{Na^2}\Theta(r_{mn} -R^{s}),~\text{otherwise}
\end{cases} 
\end{equation}
To construct $\bs{A}_{p} (\bs{r})$, we considered single unit magnetic field profile $\Big[ \frac{B_{2}}{N} + (B_{1} - B_{2})\Theta(R^{s} - r_{m,n}) \Big] \bs{\hat{z}} $ for a specific $(m,n)$. The vector potential for such field can be found in azimuthal gauge using Stoke's law. Superposing such units for all possible $(m,n)$ and doing suitable gauge transformation one gets  $\bs{A}_{p} (\bs{r})$ (for details see \cite{suppli}). 

With canonical momentum operator now defined as $\bs{\Pi} =\bs{p}+\frac{e}{c}(\bs{A}_{u}+\bs{A}_{p})$ 
, the eigenvalue equation for the hamiltonian in  $(\ref{Ham1})$, is rewritten as
\begin{gather}
 v_{F} \left[ {\begin{array}{cc}
   0 & \Pi_{x}-i \Pi_{y} \\
    \Pi_{x}+i \Pi_{y} & 0 \\
 \end{array} } \right] 
  \begin{bmatrix} \psi^{a}_{k_{x}, k_{y}} \\ \psi^{b}_{k_{x}, k_{y}} \end{bmatrix}
 =E 
 \begin{bmatrix} \psi^{a}_{k_{x}, k_{y}} \\ \psi^{b}_{k_{x}, k_{y}} \end{bmatrix}
\label{direq}\end{gather}

Here the periodic vector potential $\bs{A}_{p}$ has the 
periodicity of the square lattice magnetic modulation. Thus the magnetic translation operator $M_{\bs{R}}$ corresponding to
$\bs{\Pi}$ is same as the magnetic translation operator \cite{Zak, Fischbeck, Florek}  of an electron in 
periodic scalar potential of a square lattice and uniform transverse magnetic field. Explicitly 
\beq
M_{\bs{R}} =exp{(\frac{i}{\hbar}\bs{R}.(\bs{p}-\frac{e \bs{A}_{u}}{c}))} \label{magtran} 
 \eeq 
 We note that by construction $[\bs{p}+ \frac{e }{c}\bs{A}_{u}, M_{\bs{R}} ] =0$. Since the periodic part of the vector potential 
$\bs{A}_{p}(\bs{r} + \bs{R})=\bs{A}_{p}(\bs{r})$, we have $[\bs{A}_{p}, M_{\bs{R}}] =0$. 
 Thus the hamiltonian in (\ref{Ham1})  or in (\ref{direq}) commutes with $M_{\bs{R}}$ and 
and they have simultaneous eigenstates which are magnetic Bloch functions \cite{Zak2}. 
This result only depends on the periodicity of $\bs{A}_{p}$ and not on the explicit form (\ref{Ap}).
The significance of this 
result is that it is now possible 
to obtain the spectral properties as well as, using TKNN approach, the transport properties for the quasiparticles described by the hamiltonian (\ref{Ham1}). 

To obtain the spectrum we use standard procedure of decoupling of Eq. (\ref{direq}) and  get the eigenvalue equations for each sub-lattice. This becomes
\beq
\hat{H}_{G}(x,y) \psi_{\bs{k}}^{a,b}(x,y)=E^{2}  \psi_{\bs{k}}^{a,b}(x,y) \label{eigval}
\eeq
with 
\beq
\hat{H}_{G}  =  v_{F}^2((p+ \frac{e }{c}A_{u})^2 + V_{p}(\bs{r}) + V_{np}(\bs{r})) \label{ham} \eeq 
where 
$V_{p}(\bs{r})=\frac{e^2}{c^2}A_{p}^2 + \frac{2e}{c} \bs{A}_{p}.\bs{p} + \hbar \frac{e}{c}( B_{u} +
B_{p}) $, $V_{np}(\bs{r})  =  2\frac{e^2 }{c^2}\bs{A}_{u}.\bs{A}_{p} $
such that $V_{np}(\bs{r} +\bs{R})=V_{np}(\bs{r}) +  2\frac{e^2 }{c^2} \bs{A}_{u}(\bs{R}).\bs{A}_{p} (\bs{r})$ defines the non-periodic part of the potential.  
The hamiltonian $H_{G}$  is equivalent to the Harper-Hofstadter hamiltonian \cite{Hofstadter, Indubala}, but with the non-periodic potential $V_{np}(\bs{r})$. 

Since  the eigenfunctions are the 
magnetic Bloch functions, namely the eigenstates of $M_{\bs{R}}$ in Eq. (\ref{magtran}),
to write the eigenvalue equation (\ref{eigval}) in a tight binding form  we expand them 
in terms of localised Wannier functions in presence of
uniform transverse magnetic field $\bs{B}_{u}$\cite{Luttinger, Wannier}. 
\beq \psi_{\bs{k}}^{a} = \sum_{i} g(\bs{R}_{i}) \exp(-i \frac{e\bs{A}_{u} \cdot \bs{R}_{i}}{\hbar c}) w_{0}( \bs{r} - \bs{R}_{i}) \label{wf} \eeq
\begin{figure}[H]
\includegraphics[width=9.5cm, height=7cm]{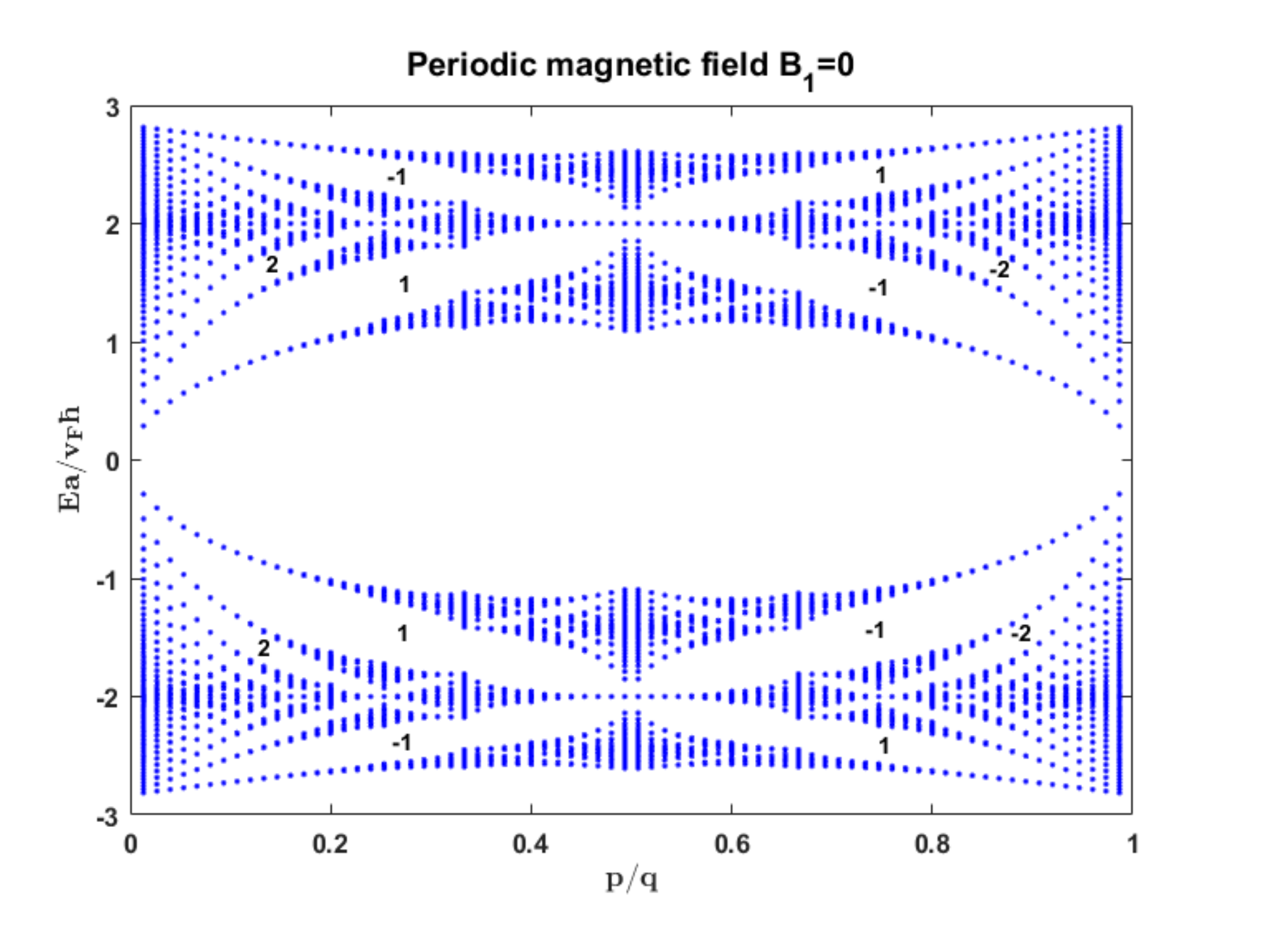}
\includegraphics[width=9.5cm, height=7cm]{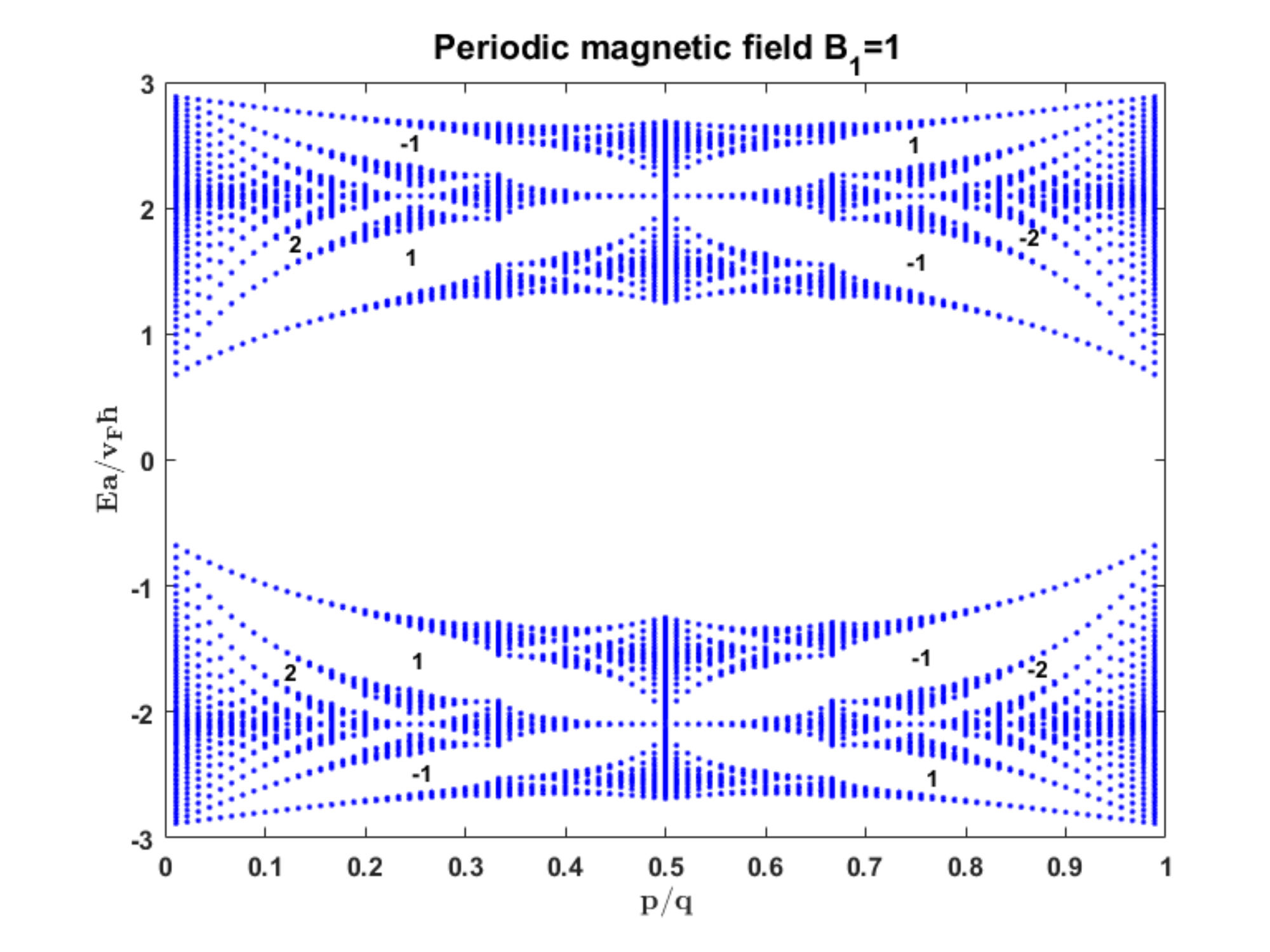}
\caption{Dimensionless energy $\frac{Ea}{\hbar v_{F}}$ versus $\frac{p}{q}$ (Eq. \ref{finalharper}) plot with Top: $B_{1}=0$ and Bottom: $B_{1}=1$.  The values of p satisfy $1 \leq p \leq q-1$, $a=0.25$ and $R^s=0.0825$ respectively.}
\label{Fig2}
\end{figure} 

\begin{figure}[h!]
\includegraphics[width=9cm, height=5.5cm]{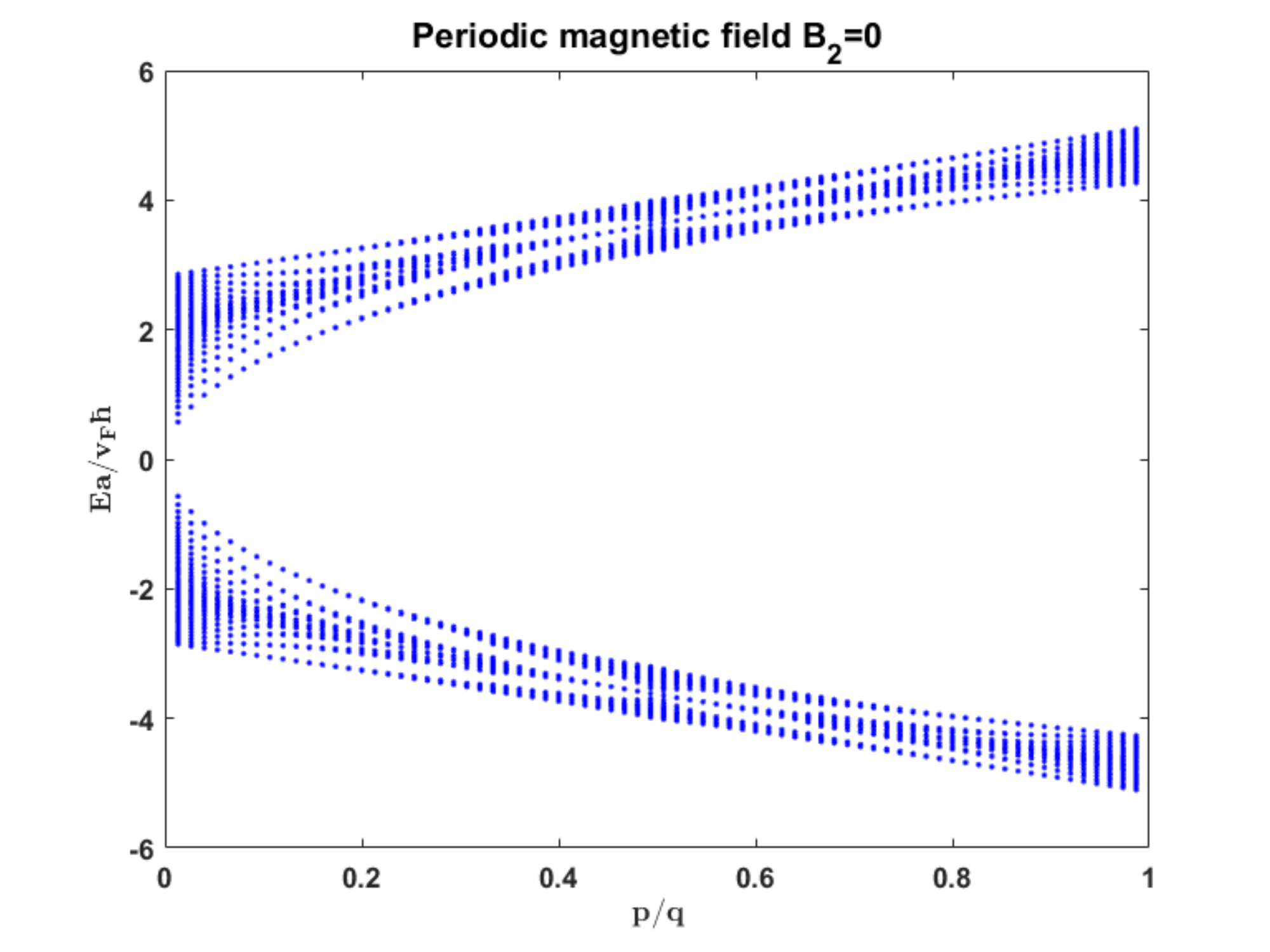}
\includegraphics[width=9cm, height=5.5cm]{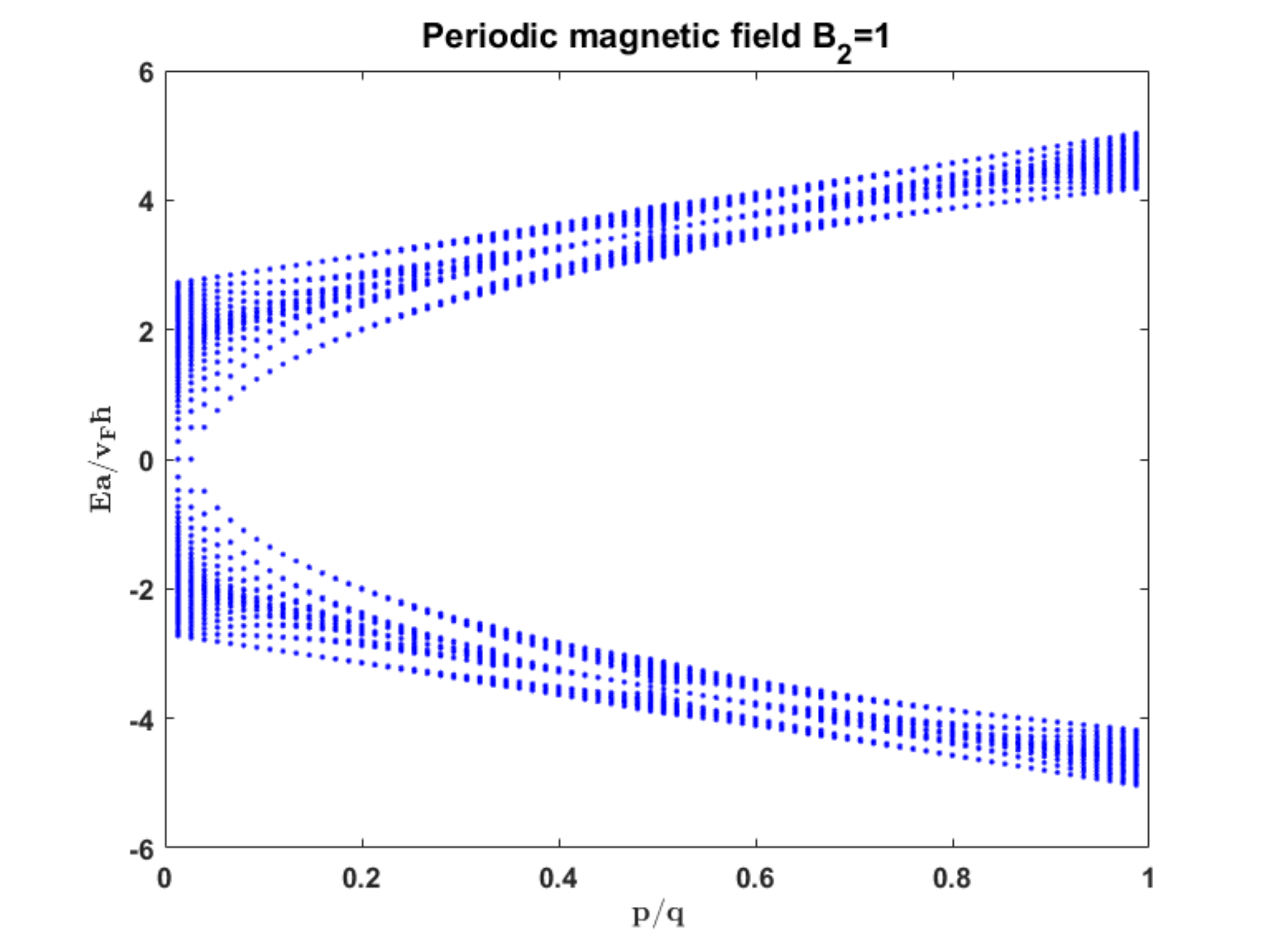}
\caption{Dimensionless energy $\frac{Ea}{\hbar v_{F}}$ versus $p/q$ (Eq. \ref{finalharper}) plot with Top: $B_{2}=0$ and, Bottom: $B_{2}=1$. Other parameters are same as Fig. $\ref{Fig2}$}
\label{Fig3}
\end{figure}
To simplify the problem further we set the condition $|\frac{ea^{2}}{\hbar c} (B_{u}+B_{p})|<<1 $. 
For $B_{p}=0$, this condition translates into $a \ll l_{B_{u}}, (l_{B_{u}}=\sqrt{\frac{\hbar c}{eB_{u}}})$, implying weak and slowly varying magnetic field \cite{Luttinger, exact}.
 
This type of condition is used in lattice gauge theory calculation  \cite{Governale}. Here using this we can write the eigenvalue equation ($\ref{eigval})$ in the form of discrete Schr\"odinger equation which takes the form of
Hofstadter-Harper equation(details in \cite{suppli}).
\begin{widetext}
\bea
\epsilon g(m,n) &=&e^{\frac{i ea}{\hbar c} (A_{ux} + A_{px})}  g(m+1,n) + e^{-\frac{i ea}{\hbar c} (A_{ux} + A_{px})}  g(m-1,n) +  e^{\frac{i ea}{\hbar c} (A_{uy} + A_{py})} g(m,n+1) \nn\\
&+& e^{-\frac{i ea}{\hbar c} (A_{uy} + A_{py})} g(m,n-1) - (\frac{e a^2}{\hbar c} (B_{u} +B_{p}) + 4) g(m,n) \label{finalharper}
\eea
\end{widetext}
with $\epsilon=\frac{-E^2 a^2}{v_{F}^2 \hbar^2 }$. 
It may be pointed out that in absolute value the magnetic field may still be substantially high and can provide substantial gap in the energy spectrum if the lattice separation 
is of the order of $\sim 100 nm$ which is possible within current technology \cite{Bader}.

The magnetic translation operator in Eq. $(\ref{magtran})$ forms magnetic translation group satisfying the algebra \cite{Brown}
\bea
M_{\bs{R}_{1}}M_{\bs{R}_{2}} & = & \exp (\frac{2 \pi i }{\phi_{0}} \phi) M_{\bs{R}_{2}} M_{\bs{R}_{1}} \nn \\ 
M_{\bs{R}_{1}} M_{\bs{R}_{2}} & = & \exp (\frac{\pi i }{\phi_{0}} \phi) M_{\bs{R}_{1}+\bs{R}_{2}}   \label{mr3} 
\eea
where 
$\phi=\bs{B}_{u}.(\bs{R}_{1} \times \bs{R}_{2}) = \frac{p}{q} \phi_{0}$ is chosen as a  rational number of flux quanta that passes through a unit cell. This 
defines the magnetic unit cell 
$\bs{R}'=m(q)a \bs{\hat{x}}+na\bs{\hat{y}}$ such that $B_{u} qa^{2} = p $, an integer. This  gives 
\beq
B_{2}=((p/q) \phi_{0}-B_{1} \pi Rs^2)/(a^2-\pi Rs^2) \label{constraint1}
\eeq
The corresponding Magnetic Brillouin zone (MBZ) is defined as $0 \le k_{x} \le \frac{2 \pi}{qa}$ and $0 \le k_{y} \le \frac{2\pi}{a}$. 
If the redefined magnetic lattice  has $N_{q}$ points along the $x$-axis and $N_{y}$ points along the $y$-axis then by construction 
$N_{q} q = 2N_{max}+1 = N_{y}$. The two ends of the lattice along $x$-axis  and $y$-axis are connected by $\bs{R}_{x}= q N_{q}a \bs{\hat{x}}$ and $\bs{R}_{y} = N_{y}a \bs{\hat{y}}$ respectively. To solve the eigenvalue problem one can use 
condition along $x$ and $y$ axes as 
$M_{\bs{R}_{x}}\psi_{\bs{k}}^{a}(x,y)) = \psi_{\bs{k}}^{a}(x,y)$ and $M_{\bs{R}_{y}}\psi_{\bs{k}}^{a}(x,y)) = \psi_{\bs{k}}^{a}(x,y)$
where $\psi_{\bs{k}}^{a}(x,y))$ are the eigenfunctions of $H_{G}$. It can be checked \cite{Brown, Munoz} that 
if the number of fluxes through each unit cell is a rational number $\frac{p}{q}$, such boundary condition can be satisfied. 


With  the relation $(\ref{constraint1})$ and the mentioned boundary conditions we numerically solve the Eq. $(\ref{finalharper})$. The results are plotted in Fig. \ref{Fig2} and Fig. \ref{Fig3}. In Fig. \ref{Fig2} (Top) and (Bottom)  keeping  $B_{1}=B_{u}+B_{p}$
fixed we plot the dimensionless energy $\pm \sqrt{-\epsilon}$  for the particle-hole spectrum \cite{Kohmoto2} as a function of $\frac{p}{q}$ for two different values of  $B_{1}$. The increase in the gap between the 
particle and hole spectrum with increasing $B_{1}$ is clearly visible. The Fig. $\ref{Fig3}$ on the other hand plots the same energy spectrum, but keeping $B_{2}$ fixed, for two representative values of $B_{2}$.
For each value of $B_{2}$, in accordance with Eq. $(\ref{constraint1})$, $B_{1}$ varies with $\frac{p}{q}$ as we move along the horizontal axis. This causes variable gap between the particle and hole spectrum as $\frac{p}{q}$ changes.
After discussing the spectrum of the hamiltonian in Eq. $(\ref{direq})$, we shall now evaluate the Hall conductance of the system and connect it with the spectrum.

The eigenfunctions of the hamiltonian (\ref{direq}) can also be written in the Bloch form  as 
\begin{gather}
 \begin{bmatrix} \psi^{a}_{k_{x}, k_{y}} \\ \psi^{b}_{k_{x}, k_{y}} \end{bmatrix}
 =
 e^{i \bs{k}.\bs{r}}
  \begin{bmatrix}
   u^{a}_{k_{x}, k_{y}} \\
   u^{b}_{k_{x}, k_{y}}
   \end{bmatrix} \label{psiu}
\end{gather}
We define 
 $u'_{k_{x}, k_{y}}= \begin{bmatrix}
         u^{a}_{k_{x}, k_{y}} \\
           u^{b}_{k_{x}, k_{y}} 
         \end{bmatrix}.$
Since $B_{u} \neq 0$, the commutation relation (\ref{mr3}) implies that 
$u^{a,b}_{k_{x}, k_{y}}$ must have zeros inside a magnetic unit cell  and have the structure $u^{a,b}_{k_{x}, k_{y}} = |u^{a,b}_{k_{x}, k_{y}}(x,y)| \exp(i \theta^{a,b}_{k_{x},k_{y}}(x,y))$ \cite{Kohmoto}. 
These functions can be found by solving the Schr\"odinger Eq. $\hat{H}(k_{x}, k_{y}) u'^{n_{b}}_{k_{x}, k_{y}} = E^{n_{b}} u'^{n_{b}}_{k_{x}, k_{y}}$, where $n_{b}$ is the band index.
Following TKNN \cite{Thouless1} the Hall conductance calculated through Kubo formula in the linear response regime for a completely filled band (the band index is omitted due to single band)
can therefore be obtained as 
\beq
\sigma_{xy}=\frac{e^2}{h} \frac{1}{2 \pi i} \int d^2 k [\bs{\nabla_{k}} \times \bs{\hat{A}}(k_{x}, k_{y}) ] _{3} \label{501}
\eeq 
where the integration is over the MBZ and 
\beq
\bs{\hat{A}} (k_{x},k_{y})=\int d^2 r u'^*_{k_{x}, k_{y}} \bs{\nabla_{k}} u'_{k_{x}, k_{y}}
\eeq
is the Berry connection defined over such MBZ. 
Since the  two points $k_{x}=0$ and $k_{x}= \frac{2 \pi}{q a}$ (or $k_{y}=0$ and $k_{y}= \frac{2 \pi}{a}$) are equivalent, the MBZ has topology
of a torus. The phase of the wave-function cannot be uniquely determined in the MBZ because of the existence of zeros of such wave function. This leads to finite value of the 
above integral and quantization of the Hall conductivity in this case giving 
\beq
\sigma_{xy}=\frac{2 e^2}{h} \sigma_{H} \label{QHE}
\eeq  
where $\sigma_{H}$ is an integer. Here the factor 2 is coming due to sub-lattice degrees of freedom in single layer graphene.
Thus if the Fermi energy lies in one of the gaps of the spectrum, 
the Hall conductivity is quantized in terms of an integer \cite{Thouless1} relevant to the filled band lying below the Fermi energy. Perturbations like disorder, interactions or effect of the higher order terms 
neglected in (\ref{finalharper}) can negate the
above result only if the gap closes. Details of the calculation closely follow 
\cite{Kohmoto} and are summarized in \cite{suppli}. 

To relate the integers  in Eq. (\ref{QHE})  with spectrum plotted in Fig. \ref{Fig2}, we note that integration of 
the Berry connection  $\bs{\hat{A}} (k_{x},k_{y})$  gives the geometric phase change of the functions $u'_{k_{x}, k_{y}}$ over 
the MBZ. Following \cite{Dana, Wilkinson} it is possible to show that the corresponding magnetic Bloch functions satisfy the boundary condition \cite{suppli}
\beq
\psi_{k_{x}+\frac{b_{1}}{q},k_{y}}=\psi_{k_{x},k_{y}} , ~\psi_{k_{x},k_{y}+b_{2}}=e^{i \sigma_{H} k_{x} qa} \psi_{k_{x},k_{y}} \label{bcf} 
\eeq
where $\bs{b}_{1}=\frac{2 \pi}{a} \bs{\hat{x}}$ and $\bs{b}_{2}=\frac{2 \pi}{a} \bs{\hat{y}}$ are the reciprocal lattice vectors. 
This condition must be consistent with the group properties defined in Eqs. (\ref{mr3})
of the magnetic translation operator, given in Eq. (\ref{magtran}), leading to  
\beq
M_{a\bs{\hat{y}}}M_{a\bs{\hat{x}}} \psi_{\bs{k}}=e^{i (\bs{k}+ \frac{p}{q} \bs{b}_{2} ). a\bs{\hat{y}}}  M_{a\bs{\hat{x}}} \psi_{\bs{k}} \label{comm}
\eeq
where $M(a \bs{\hat{x}}) \psi_{\bs{k}}$ depicts eigenfunction of $M(a \bs{\hat{y}})$ with eigenvalue $e^{i (\bs{k}+ \frac{p}{q} \bs{b}_{2} ).a \bs{\hat{y}}}$ . 
In the same way $M_{a \bs{\hat{x}}} \psi_{\bs{k}}$, ....$M_{(q-1)a \bs{\hat{x}}} \psi_{\bs{k}}$ all have different eigenvalue for $M_{a \bs{\hat{y}}}$, but same eigenvalue for the Hamiltonian. This leads to a $q$ fold degeneracy for each energy eigenvalue.
The algebraic relation that connects the boundary condition in Eq. (\ref{bcf}) with the relation (\ref{comm}),  connects $\frac{p}{q}$ 
to the integer $\sigma_{H}$ through the famous Diaphantine Equation 
\beq
\mu q +\sigma_{H}p=1 \label{Dio1}
\eeq
This equation is equivalent to the equation given by TKNN \cite{Thouless1}, for square lattice systems to calculate the Chern numbers associated with $\sigma_{H}$  for $r$th gap in a Landau level as 
$r=s_{r}q+t_{r}p$. Under the constraint $|t_{r}|\leq q/2$ this yields a unique solution $(s_{r}, t_{r})$. Here $\sigma_{H}=t_{r}-t_{r-1}$ and $\mu=s_{r}-s_{r-1}$ \cite{Avron}. 
In our system with square lattice magnetic modulation using the same method we calculate $t_{r}$ for few such swaths  in Fig. \ref{Fig2}. 

To summarise we predict topological quantization of Hall conductivity for massless dirac fermions in monolayer graphene 
under generic two dimensional periodic magnetic modulation. The results can be extended to a non-relativistic 2DEG. 
Similar topological quantization in non bravais type of magnetic modulation in hexagonal lattice \cite{Rammal},  effect of additional periodic electrostatic potential and (in)commensuration between two types of periodicities \cite{Hofstadter, Gerhardts, Gumbs}, bulk and edge  correspondence in such system\cite{Halperin, Muller} will be some possible directions for further investigations.  
Periodic two dimensional magnetic modulation was already created for 2DEG in GaAs-AlGaAs heterojunctions using ferromagnetic 
dysprosium dots \cite{expt1}. Van der Waals heterostructure of monolayer graphene \cite{expt2} with layered magnetic materials such as Transition metal phosphorus trisulphide \cite{expt3} may also realize similar modulations. 

We acknowledge helpful discussion and correspondence with S. Gupta, S. Shringarpure, M. Sharma and Itzhack Dana. MA is supported by a MHRD fellowship.

\newpage
\def \bb{\bibitem}
\def \bb{\bibitem}

\end{document}